# On fractionally linear functions over a finite field

V.M.Sidelnikov[1], R.N.Mohan[2] and Moon Ho Lee[3]
University of Moscow, Russia[1],
Sir CRR Institute of Mathematics[2], Eluru-534007, AP, India
Institute of Information & Communication[3], Chonbuk National University, Korea
Emails: sidd123@hotmail.com[1], mohan420914@yahoo.com[2], moonho@chonbuk.ac.kr[3],

**Abstract:** In this note, by considering fractionally linear functions over a finite field and consequently developing an abstract sequence, we study some of its properties.

## 1. Introduction

The fractionally linear function, refer to [6, 7], means it is equal to the ratio of two linear functions like $f(z) = \dfrac{az+b}{cz+d}$, where z is a variable possessing values in some finite field F and elements a, b, c, d are constants from the field F such that ad ≠ bc, i.e. $\begin{vmatrix} a & b \\ c & d \end{vmatrix} \neq 0$. Then the function $f(z)$ is called Mobius transformation or linear fractional function or transformation. For some details of the field under consideration refer to Sidelnikov [8].

In this paper we consider the group of all fractional linear transformations of the complex plane with coefficients (some constants) either in □ or in □. For detailed study of groups refer to Rotman [5]. This is essentially the same, up to a factor of 1, as SL ( F ), which is the group of 2-by-2 matrices with entries in □ and determinant 1. This SL (Z) is called as the modular group.

A special type of automorphic function, where the group involved is the modular group is being considered. Usually this term is applied in a more general way. We can define $\varphi(x)$ as a function where v is an integer. Then the condition may be applied for certain subgroups of finite index in the modular group. We denote the group of fractionally linear functions by $FL(F_q)$, where $F = F_q$, the finite field with q elements and the order of $FL(F_q)$ is equal to $q(q^2-1)$. Note that $SL(F_q)$ and $FL(F_q)$ are different groups, which are at the same time isomorphic groups denoted by $SL(F_q) \cong FL(F_q)$.

Now by considering such modular groups with automorphism, we develop a sequence and define correlation functions on that, and conclude by suggesting an application for the systems that are being developed in this paper.

## 2. Main results

Let $\psi \in FL(F_q)$ be the fractionally linear function over the finite field $F_q$. Suppose that $\psi$ has no fixed point in $F_q$ and the order of $\psi$ as the element $FL(F_q)$ is equal to q+1. In this case the set $\{\alpha_0, \alpha_1, \ldots, \alpha_q\} = F_q \cup \{\infty\}$ can be represented in the following form:

$$F_q \cup \{\infty\} = \{\psi^j(1) \mid j=0,\ldots,q\}$$

Let $\varphi$ be some function of group $FL(F_q)$. Then we can define the sequence $a_\varphi$ as follows:
$$a_\varphi = (\varphi(\psi^0(1)), \varphi(\psi^1(1)), ...., \varphi(\psi^q(1))). \qquad (1)$$
For further details of these sequences refer to [2, 6]. Then the cyclic shift $a_\varphi^{(s)}$ on s co-ordinates of sequence $a_\varphi$ can be written in the form

$$a_\varphi^{(s)} = (\varphi(\psi^s(1)), \varphi(\psi^{s+1}(1)), ..., \varphi(\psi^{s+q}(1))).$$

Let $\chi(x)$ be a character of the multiplicative group of the finite field $F_q$ with q elements. The finite fields have been studied by Sidelnikov [7]. Let us assume that $\chi(0) = \chi(\infty) = 1$. Then consider the above mentioned sequence with this character, which has the following form

$$a_\varphi^\chi = (\chi(\varphi(\psi^0(1))), \chi(\varphi(\psi^1(1))), ....., \chi(\varphi(\psi^q(1)))),$$

with entries from $\square$.

The sequence $a_\varphi^{\chi,(s)}$ is the cyclic shift to the left of the sequence $a_\varphi^\chi$ on s coordinates and it can be represented in the following form

$$a_q^{\chi(s)} = (\chi(\varphi(\psi^s(1))), \chi(\varphi(\psi^{s+1}(1))), ..., \chi(\varphi(\psi^{s+q}(1)))).$$

The periodic auto correlation function of sequence $a_\varphi^\chi$ can be defined as

$$T_s(a_\varphi^\chi) = \langle a_\varphi^\chi, a_\varphi^{\chi,(s)} \rangle, \text{ where } s = 0, ..., q,$$

and $\langle .. \rangle$ denotes the inner product in the unitary space $\square$. For further details regarding sequences refer to [3].

Now consider the sequence $a_\varphi^\chi = (\chi(\varphi(\psi^0(1))), \chi(\varphi(\psi^1(1))), ....., \chi(\varphi(\psi^q(1))))$,

then the cross correlation function of the sequences $a_\varphi$ and $a_{\varphi'}$, can be defined as

$$T_s(a_\varphi, a_{\varphi'}) = \langle a_\varphi^\chi, a_{\varphi'}^{\chi,s} \rangle, \text{ where } s = 0, ..., q.$$

The maximum non-trivial correlation $T_{max}$ of the sequence $a_\varphi^\chi$ is defined as

$$T_{max}(\varphi) = \max_{0 < s \le q} |T_s(a_\varphi^\chi)|$$

The maximum non-trivial cross correlation of sequences $a_\varphi^\chi$ and $a_{\varphi'}^\chi$ is

$$T_{\max}(a_\varphi, a_{\varphi'}) = \max_{0 \leq s \leq q} \left| T_s(a_\varphi^\chi, a_{\varphi'}^\chi) \right|,$$

In order to facilitate synchronization as well as to minimize interference due to the co-existence of other sequences, the maximal non-trivial auto-and cross-correlation values of this sequence family should be as small as possible. Welch bound, Sidelnikov bound and Sarwate bound, (for further details refer to Levenshtain [4]), provide lower bounds on the minimum possible value of the parameter $T_{\max}$ and commonly used to judge the merits of a particular sequence design. Now we will discuss them.

By using the Sarwate's result, we have

$$\frac{|a_\varphi|^2}{N} = \frac{N-1}{N(M-1)} \frac{|a_{\varphi'}|^2}{N} \geq 1,$$

where $a_\varphi = (a_{\varphi_{1,i}}, ..., a_{\varphi_{q,i}})$ and $|a_\varphi| = \left( \sum_{i=0}^{q} |a_{\varphi,i}|^2 \right)^{1/2}$ is Euclidean norm or energy of the sequence and $N = q+1$.

This Welch bound applies to complex-valued sequences in general, while Sidelnikov bound, applies only to those sequences whose symbols are complex $q^{th}$ roots of unity. In most of the cases, this Sidelnikov bound is tighter than that of Welch bound, restricting the symbols to the complex roots of unity. When $M = N^u$, and $N \Box u$ and $u \geq 1$ is an integer, the Sidelnikov bounds can be well approximated by:

$$M_A(n,d) \leq \frac{2n^3 - 2n(n-2d)^2}{3n - (n-2d)^2 - 2}, \quad ...(2)$$

where $M_A(n,d)$ is the number of elements of antipodal binary code.

Consider the remarkable binary Kerdok code $K$ that has following parameters: The length $n = 2^m$, where m is even, the code distance is equal to $d = \frac{n - \sqrt{n}}{2}$ and the number of elements of $K$ is equal to $n^2$. This Kerdok code meets the bound (2).

Let $A = \{a_{\varphi_1}, ..., a_{\varphi_M}\}$ be the set of sequences, which determines by sequences $\varphi_1, ..., \varphi_M$ by means of identity (1). The value of cross correlation of sequences from the set $A$ is defined by

$$T(A) = \underset{\substack{1 \leq i \leq j \leq M \\ s \neq 0 \; if \; i=j}}{Max} \left| T_s(a_{\varphi_i}, a_{\varphi_j}) \right|$$

Then in the considered case the Welch bound has the following form

$$T_{\max}(A) \geq N\sqrt{\frac{M-1}{NM-1}}$$

Then we have the following

$$T_{\max}^2 \geq \begin{cases} N\left(2u+1-\dfrac{1}{1.3.5...(2u-1)}\right) \text{ for binary sequences} \\ N\left(u+1-\dfrac{1}{1.2.3..u}\right) \text{ for non binary sequences of the present type} \end{cases}$$

These approximations may improve the performance of the sequences and when $M \sqcup N$ then consequently $u \sqcup 1$, (this $\sqcup$ indicates approximation). Then we can say that

$$T_{\max} \geq \begin{cases} \sqrt{2N} \text{ for binary sequences} \\ \sqrt{N} \text{ for non binary sequences of the present type.} \end{cases}$$

Let $a$ be sequence of the length N with entries in $F_q$ i.e. $a \in F_q^N$. The linear span L($a$), of $a$ is the length of the shortest linear recursion over the field $F_q$ under consideration, which is generated sequence $a$. The linear span of a sequence is one measure of its predictability. Any "good" pseudo-random sequence should have large linear span compared with its length. If a sequence has little linear span then this sequence has predictability on its initial section.

If a sequence has a linear span $L = \ell$, then its linear recursion can be determined from any $2\ell$ successive elements of this sequence by means of Berlekamp algorithm, for details refer to [1,2], which provides an elegant way to factor polynomials over a small finite field of order q. The remaining elements of the sequence can be produced from the recursion.

For an application point of view, refer to [6], these systems are being extensively used for control systems, besides by using Fuzzy logic and Fuzzy control systems.

**Acknowledgements:** This work was supported by the MIC (Ministry of Information and Communication), under the ITFSIP (IT Foreign Specialist Inviting Program ) supervised by IITA, under ITRC supervised by IITA, and International Cooperation Research Program of the Ministry of Science & Technology, Chonbuk National University, Korea. So we convey our thanks to all the concerned.